\documentclass[12pt]{iopart}
\usepackage{graphicx}

\begin{document}
\newcommand{\dif}{\ensuremath{{\rm d}}}
\title[Semi-analytical black body radiation]{A semi-analytical approach
  to black body radiation} 
\author{C Calcaneo-Roldan$^1$,  O Salcido$^2$, D Santana$^{1,2}$}
\address{$ˆ1$ Departamento de F\'isica, Universidad de Sonora, Rosales y
  Blvd Luis Encinas S/N Col Centro, Hermosillo, Sonora 83000, M\'exico}
\address{$ˆ2$ Departamento de Investigaci\'on en  F\'isica, Universidad de
  Sonora, Rosales y Blvd Luis Encinas S/N Col Centro, Hermosillo,
  Sonora 83000, M\'exico}
\ead{carlos.calcaneo@fisica.uson.mx}

\begin{abstract}
  We describe a semi-analytical method to calculate the total radiance
  received form a black body, between two frequencies. {As has been
  done before, the method takes advantage of the fact that the
  solution simplifies with the use of polylogarithm functions.} We then
  use it to study the amount of radiation from the sun received by
  bodies at Earths surface. 
\end{abstract}
\noindent{\it Keywords\/}: Black body radiation, Available energy,
Polylogarithm

\submitto{\EJP}
\maketitle

\section{Introduction}
\label{intro}

What follows started originally as a question in a second year
undergraduate course on Modern Physics. The question was simple: We
know the Boltzmann formula for the total radiation of a black body, but
what is total radiance {\it between two particular wavelengths}. It
was posed by the instructor as an assignment, but the students took it
to heart, completing a long and interesting path researching what they
now know to be Planck integrals and the Polylogarithm functions. Our
intention here is to highlight the main points learned, in an effort
to make it accessible to other undergraduate students.

A black body is an ideal theoretical object which 
absorbs all the energy that falls on its surface as light. There is no
reflection nor light passing through the black body. Due to the nature
of radiation absorption, which can be modelled as the resonance of charge
at the surface of the body, black bodies can actually emit all the
energy as light, given enough time, again in all wave lengths or
frequencies (see e.g. \cite{krane}).

Perfect black bodies seldom occur naturally, even black smoke
reflects up to $1\%$ of the incident light. The name {\it black body}
was first introduced by Gustav  Kirchhoff in the 1860's. The emitted light
of one of these objects is called black body radiation; despite its
name, it constitutes a physical model for studying the emission of
electromagnetic radiation of solids or liquids.

This radiation refers, therefore, to the continuous energy emission
from the surface of any body, which is transported by
means of electromagnetic waves that can travel in the vacuum at the speed of 
light. The radiant energy emitted  from a body at room temperature is
small and corresponds to wavelengths greater than those of visible
light (lower frequency). With increasing temperature the emitted
energy increases and the wavelengths become shorter. This is the
reason we see a change in the colour of a body when heated. Bodies
do not emit with the same intensity in all frequencies or wavelengths,
a good model for this intensity is Planck's law for black body
radiation, which is a function of the body's
temperature\cite{crepeau}. The emitted energy of a solid also depends
on the 
nature of its surface; we can then have a matte or black surface with
larger emission power than that of a glossy one. The energy emitted
by an incandescent charcoal filament is larger than that of a platinum
filament at the same temperature. { Therefore, real bodies will emit 
radiation with varying efficiency (a concept known as emissivity),
in these cases the well known black body spectra must be adapted to
give more realistic answers (see e.g. Ram\'irez-Moreno} {\it et al.}
\cite{ramirez15}, {and references therein).}
The surface of a black body is the
limiting case in which all the light from the exterior is absorbed and
all the interior energy is emitted\cite{poprawski}.  

We may find the theoretical model for black body radiation in any
introductory text on Modern Physics. The expression for the energy of
the radiation in an interval ${\rm d}\nu$ is a rather simple
relation and it turns out that this density is proportional to the total
radiance emitted by the black body\cite{zombeck}. However, if we are
to calculate the total energy between two frequencies $\nu_1$ and
$\nu_2$ we must use special functions or implement some numerical
technique.

{The importance of radiation absorption and emission has lead to the
study of black body radiation for many years and in many fields.
A report on an exact solution, written as an infinite series, can be found
in the work by S.L. Chang and K.T. Rhee}\cite{CHANG1984451}.{ A very
complete discussion of how this solution may be rewritten in terms of
polylogarithms is discussed by Se\'an
M. Stewart}\cite{Stewart2012232}. Although developed in an
independent way, we will build on these ideas and implement the
functions numerically for particular ranges in wavelength. {The subject
of integrating black body radiation in a particular range of
wavelength is also known as band emission and has been extensively
discussed by the infrared imaging community (see e.g. Vollmer and
M\"ollmann}\cite{vollmerbook} {and references therein).}

In this work We take data from recent radiation measurements and aim
to implement a semi-analytical approach for calculating the fraction of
radiation from the Sun that is available at ground level. These
estimates may be important for recent applications in renewable energy
sources for calculating the total radiance available to solar panels
and may be easily adapted for each region for which the radiance map
is known (or may be measured).

The paper is organised as follows: after this brief introduction,
section \ref{model} presents the semi-analytical technique, in section
\ref{example} we apply it to some generic data and our results are
summarised in the conclusions. 

\section{Energy density between two frequencies.}
\label{model}

The full description for black body radiation was not realised until
1900 when, almost reluctantly, Max Planck was able to construct an
expression that reproduced all of the data (unlike the previous intents
by Lord Rayleigh and Sir James Jeans, for low frequencies and Wien and
Planck for higher frequencies\cite{kragh}) .  

\begin{center}
  \begin{figure}
    \centerline{\includegraphics[width=.35\textwidth]{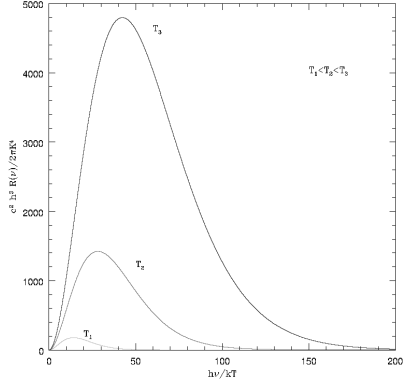}}
    \caption{Spectral radiance for a body at three temperature values,
      $T_1<T_2<T_3$. As| temperature increases, the maximum of the curve
      moves to higher frequencies and the radiance is significant for a
      greater interval of frequencies.}
    \label{fig:rad}       
  \end{figure}
\end{center}

The details of these developments can be found in many textbooks on
modern physics (eg, \cite{krane,eisberg1}). The description centres on
considering a cavity with metallic walls heated uniformly to
temperature $T$. The walls emit electromagnetic radiation in the
thermal range of frequencies. Rayleigh and Jeans used classic
electromagnetic theory to demonstrate that the radiation in the
interior of the cavity must exists as standing waves whose nodes
correspond to the metallic surfaces. In the classic theory, the
average of total energy only depends on temperature $T$. The number of
standing waves in the frequency interval times the average energy of
these waves, divided by the volume  
of the cavity, gives the average energy content per unit volume in the
frequency interval between $\nu$ and $\nu+{\rm d}\nu$. This is the
required quantity: the energy density $\rho_T(\nu)$. By considering
that these standing waves contained in the cavity must have discrete
energy values, we may write the expression for the spectral energy density: 
$$\rho(\nu){\rm d}\nu = \frac{8\pi{\rm h}}{\rm c^3}\frac{\nu^3}{{\rm
  e}^{\frac{\rm h\nu}{{\rm k}T}}-1}{\rm d}\nu$$
and then obtain the {\it spectral radiance} of the body as
a multiple of this density\cite{zombeck}:  
\begin{equation}
R(\nu){\rm d}\nu=\frac{c}{4}\rho(\nu){\rm d}\nu
\end{equation}
This quantity is defined as the energy radiated per unit length per
unit time in a frequency interval ${\rm d}\nu$. We illustrate this radiance
for three distinct temperature values $T_1<T_2<T_3$ in figure
\ref{fig:rad} (throughout this paper we use the usual notation for the
Planck constant, ${\rm h}$, the speed of light in the vacuum, ${\rm
  c}$, and the Boltzmann constant, ${\rm k}$).

\subsection{Energy density and polylogarithms}

Once we know the expression for the spectral energy density of a black
body it is not difficult to integrate the total energy per unit volume
available between two specific frequencies, $\nu_a$ and $\nu_b$:
\begin{equation}
 \rho_{ab}(T) \equiv
 \int_{\nu_a}^{\nu_a}\rho(\nu){\rm d}\nu=\int_{\nu_a}^{\nu_b}\frac{\rm
   8\pi h}{c^{3}}\frac{\nu^{3}}{{\rm e}^\frac{\rm h\nu}{{\rm k}T}-1}{\rm d}\nu
 \label{radab}
\end{equation}
Since the two frequencies between-which we would like to calculate
the energy density are constant, the quantity that we have just  
defined in equation (\ref{radab}) is only a function of the
temperature of the black body (i.e. we will treat $\nu_a$ and $\nu_b$
as parameters). Of particular importance and historical
significance is $ \rho_{0\infty}(T) \equiv
\rho_T(T)$ which is the {\it total energy density of the black body},
integrated over all frequencies, we will get back to this integral
as a specific case of the the general expression between two frequencies.

In order to handle the integral more easily, we suggest using the
following adimensional variables :
  \[\begin{array}{ccccc}
  \chi=\frac{\rm h\nu}{{\rm k}T} & & {\rm and} & &  \rho
  (\chi){\rm d}\chi = \frac{8\pi{{\rm k}^4T^4}}{\rm c^3h^3}
  \tilde{\rho} (\chi){\rm d}\chi
  \end{array}\]
using these definitions, we can calculate the total energy density of
the black body between two frequencies, $\nu_a$ and $\nu_b$.

Before we continue, as we shall soon encounter, the spectral range
for radiation is commonly given in wavelengths, rather than
frequencies. We have written \ref{radab} because this form will turn
out to be useful for our analytical manipulation. Let us consider for
a moment that, instead of a frequency range, we have a wavelength
range, which corresponds to the same physical range for the radiation;
in this case the energy density of interest is
\[
\rho_{ab}(T) = \int_{\lambda_a}^{\lambda_b} \rho(\lambda){\rm d}\lambda
\]
which we would use if the spectral density is a function of
wavelength. Now, let us examine this integral, we could split this
problem into a couple of integrals:
\[
\rho_{ab}(T) = \int_{o}^{\lambda_b} \rho(\lambda){\rm d}\lambda -\int_{o}^{\lambda_a} \rho(\lambda){\rm d}\lambda
\]
if we further consider the specific form for $\rho(\lambda)$, and
change from wavelength to frequency, it turns out that\cite{michels}
\[\int_{o}^{\lambda_o} \rho(\lambda){\rm d}\lambda =
\int_{\nu_o}^{\infty}\tilde{\rho}(\chi){\rm d}\chi\]
where the specific frequency, $\nu_o$, and wavelength, $\lambda_o$, are
related, in the usual way, through the speed of light in the vacuum:
$c=\nu_o\lambda_o$.

In this way we may do the integral over a wavelength range by using the
equivalent frequency, and again splitting the problem into a couple of
integrals: 
\[
\rho_{ab}(T) = \beta \left[
  \int_{\chi_b}^{\infty}
    \tilde{\rho}(\chi){\rm d}\chi - \int_{\chi_a}^{\infty}
    \tilde{\rho}(\chi){\rm d}\chi 
  \right],
\]
notice that all the physical units have been accounted for in the
temperature parameter $\beta=\frac{8\pi {\rm k}^{4}T^{4}}{\rm
  c^{3}h^{3}}$. If we define the following auxiliary function: 
  \begin{equation} 
    F(\chi)=\int_{\chi}^{\infty} \frac{\chi'^3}{{\rm e}^{\chi'}-1}{\rm d}\chi' 
    \label{funcA}
  \end{equation}
  the required energy density is then:
  \begin{equation}
    \rho_{ab}(T) = \beta\left[F(\chi_b)-F(\chi_a)\right]
    \label{dens1}
  \end{equation}
  with the frequencies written as $\chi_{i}=\frac{\rm h\nu_{i}}{{\rm k}T}$
   
  To solve $F(\chi)$ we study the integrand and notice that it looks
  like the convergence of a geometric series; this leads us to
  factorise ${\rm e}^{\chi}$ and write the integrand as:
  
  \[\frac{\chi^{3}}{{\rm e}^{\chi}-1}=\chi^{3}{\rm e}^{-\chi}\cdot
  \left[\frac{1}{1-{\rm e}^{-\chi}}\right]\] 

  In this last expression we recognise the geometrical series as the
  value between brackets. Because we will consider only values $\nu>
  0$, for which $\chi>0$, we can then rewrite the integrand as
  follows:
    
  \[\frac{\chi^{3}}{{\rm e}^{\chi}-1}=\sum_{n=0}^{\infty}\chi^{3}{\rm e}^{-(n+1)\chi}\]
  
  The function $F(\chi)$ then becomes:
    \[
      F(\chi)=\sum_{n=0}^{\infty}\int_{\chi}^{\infty}\chi'^{3}{\rm
        e}^{-(n+1)\chi'}{\rm d}\chi'
    \]

  We  integrate the right hand
  side of this last expression: first we integrate by parts three
  times to reduce the exponent of $\chi$, then we evaluate the
  resulting expressions, noticing that many terms vanish for $\chi
  \rightarrow \infty$. After cleaning up the expression we can write
  the result as follows:
  
  \[F(\chi)=\sum_{n=0}^{\infty} \frac{{\rm e}^{-(n+1)\chi}}{n+1}
  \left[ \chi^3 +\frac{3\chi^2}{(n+1)}+\frac{6\chi}{(n+1)^2}
    +\frac{6}{(n+1)^3} \right] 
  \]
  In this last expression, we can change the index, $
    (n+1)\rightarrow n$, so that $F$ is written in its most compact
  form:
  
  \begin{equation}
    F(\chi)=\sum_{n=1}^{\infty} \frac{{\rm e}^{-n\chi}}{n}
    \left[ \chi^3 +\frac{3\chi^2}{n}+\frac{6\chi}{n^2}
      +\frac{6}{n^3} \right]
    \label{funcB}
  \end{equation}

  To prove convergence of \ref{funcB} we use the ratio test, which
  guarantees that a series $\sum_{n=1}^{\infty}S_n$ converges when it
  meets the following condition   
  \[L=\lim_{n\rightarrow\infty}\vert\frac{S_{n+1}}{S_{n}}\vert <1\]
  For this particular case we have:
  \[\frac{S_{n+1}}{S_n}=\frac{\frac{{\rm
        e}^{(n+1)\chi}}{n+1}}{\frac{{\rm e}^{-n\chi}}{n}}\left[\frac{\chi^3+\frac{3\chi^2}{n+1}+\frac{6\chi}{(n+1)^2}+\frac{6}{(n+1)^3}}{\chi^3+\frac{3\chi^2}{n}+\frac{6\chi}{n^2}+\frac{6}{n^3}}\right]\]  
rearranging terms and taking the limit 
\[L=
\lim_{n\rightarrow\infty}\frac{n{\rm e}^{-(n+1)\chi}}{(n+1){\rm e}^{-n\chi}}\left[\frac{\chi^{3}+\frac{3\chi^{2}}{n+1}+\frac{6\chi}{(n+1)^{2}}+\frac{6}{(n+1)^{3}}}{\chi+\frac{3\chi^{2}}{n}+\frac{6\chi}{n^{2}}+\frac{6}{n^{3}}}\right]
=\frac{1}{{\rm e}^{\chi}}
\]

Because we are only interested in cases where  $\nu > 0$ and therefore
$\chi=\frac{\rm h\nu}{{\rm k}T}>0$, $L<1$ and the series converges for all
possible values in which we will use our auxiliary function. Thus, we
may now write the integrated energy density as: 

\begin{equation}
  \int_{\nu}^{\infty}\rho{(\nu')}d\nu'
  = \beta F(\chi)
  = \beta\sum_{n=1}^{\infty}\frac{{\rm
      e}^{-n\chi}}{n}\left[\chi^{3}+\frac{3\chi^{2}}{n}+\frac{\chi}{n^{2}}+\frac{6}{n^{3}}\right]  
  \label{Ffinal}
\end{equation}

For any particular frequency $\nu$, written as
$\chi=\frac{h\nu}{kT}$, and temperature, $\beta=\frac{8\pi
  k^{4}T^{4}}{c^{3}h^{3}}$.  

To recover the the energy density we need to evaluate $F(\chi)$. For
this purpose we must quantify the following four sums:

\[
\begin{array}{ccc}
  F_{1}(\chi)=\chi^{3}\sum_{n=1}^{\infty}\frac{({\rm
      e}^{-\chi})^{n}}{n} 
  &
  \;\;\;\;\;
  &
  F_{2}(\chi)=3\chi^{2}\sum_{n=1}^{\infty}\frac{({\rm
      e}^{-\chi})^{n}}{n^{2}} \\
  \\
  F_{3}(\chi)=6\chi\sum_{n=1}^{\infty}\frac{({\rm
      e}^{-\chi})^{n}}{n^{3}}
  &
  \;\;\;\;\;
  &
  F_{4}(\chi)=6\sum_{n=1}^{\infty}\frac{({\rm e}^{-\chi})^{n}}{n^{4}} 
\end{array}
\]

In the first sum, we identify the logarithm series:
\[\ln(1-Y)=(-1)\sum_{(n=1)}^{\infty}\frac{Y^{n}}{n} \;\;\;\;
\mbox{with}\;\;\;\;  |Y|<1 \] 
Identifying ${\rm e}^{-\chi}=Y$, because we will always consider cases
such that $\chi>0$,  $|Y|<1$ and the first sum can be written as:
\[F_{1}(\chi)=-\chi^{3}\ln(1-{\rm e}^{-\chi})\]

To solve the remaining sums, we identify a special function that
arises in Quantum Field Theory, the {\it polylogarithm}, ${\rm Li}_m$ (see
e.g. \cite{Frellesvig,Vollinga} and references therein):
\[{\rm
  Li}_m(z)=\frac{(-1)^{m-1}}{(m-2)!}\int_{0}^{1}\frac{\ln^{m-2}(t)\ln(1-tz)}{t}dt=\sum_{n=1}^{\infty}\frac{z^{n}}{n^{m}} \]
Here we will use these functions, known as Nielsen's Generalised
Polylogarithms\cite{kolbig}, which allow us to reduce the problem to
a family of integrals. By identifying $z={\rm e}^{-\chi}$, we see that
each of the three remaining sums may be related to a specific
polylogarithm: 

\begin{equation}
\begin{array}{ccccc}
  F_2(\chi)&=&3\chi^2{\rm Li}_2({\rm e}^{-\chi}) & = &
  -3\chi^2{\displaystyle\int}_{0}^{1}\frac{\ln(1-t{\rm e}^{-\chi})}{t}dt \\
  && \\
  F_3(\chi)&=&6\chi{\rm Li}_3({\rm e}^{-\chi}) & = &
  6\chi^2{\displaystyle\int}_{0}^{1}\frac{\ln(t)\ln(1-t{\rm e}^{-\chi})}{t}dt \\
  &&\\
  F_4(\chi)&=&6{\rm Li}_4({\rm e}^{-\chi}) & = &
  -3{\displaystyle\int}_{0}^{1}\frac{\ln^2(t)\ln(1-t{\rm e}^{-\chi})}{t}dt
\end{array}
\label{numerics}
\end{equation}

This is as far as we can go analytically. The possible values for
$\chi$ guarantee that all of the integrals in (\ref{numerics}) will
converge to a specific value, once we have a range of frequencies. All
that remains is to apply a numerical algorithm and we may solve many
interesting problems. Two particular cases are discussed in the next
section. 

\section{Examples}
\label{example}

With the simple relation in \ref{Ffinal} we can now recover physical
interesting quantities very easily. We now consider two cases for
which the formulation developed leads to quick
results. The first is a more analytical example which exemplifies the
fundamental nature of black body radiation, while the second is a more
applied case for which integrating the parts of the spectrum may be
interesting.

\subsection{The fundamental constants of black body radiation}

Consider the relation between spectral radiance and energy
density:  
G
\[R(\nu)d\nu=\frac{c}{4}\rho(\nu)d\nu \]
Due to the fact that we can measure radiance readily, this quantity is
of interest in many fields. Using our auxiliary function, the total
radiance, $R_T$, of a black body can be calculated from it: 
\[R_T=\int_{0}^{\infty}R(\nu)d\nu=\frac{c}{4}\int_{0}^{\infty}\rho(\nu)d\nu \]
but, we also know that:
\[R_T=\sigma T^{4}\]
then
\[R_T=\frac{c}{4}\int_{0}^{\infty}\rho(\nu)d\nu=\sigma T^{4}\]

Using our expression for $F$ we can solve the middle integral easily,
\[R_T=\frac{c}{4}F(0)=\sigma T^{4}\]
We see from (\ref{Ffinal}) that $F(0)$ is a constant, in fact:
\[F(0) = \beta\sum_{n=1}^{\infty} \frac{6}{n^4} = 
    6\beta\sum_{n=1}^{\infty} n^{-4} =6\beta\zeta(4) \]
in the last step, we have used the Riemann zeta function, defined by:
\[\zeta(s) = \sum_{n=1}^{\infty} n^{-s}, \ \ \ \ \ \ \ s>1\]
which converges, and for the case ${\rm s=4}$, has a value of (see,
{\it e.g.} \cite{arfken}):
\[\zeta(4) = \frac{\pi^4}{90} \]
then ${\rm F(0)=\beta\frac{\pi^4}{15}}$ and the total energy density
can be written: 
\[\rho_T =  \frac{8\pi^5 {\rm k}^4T^4}{15{\rm c^3h^3}}\]

We have found an expression for the total radiance of a black
body.
\[R_T = \frac{c}{4}\rho_T=\frac{2\pi^5{\rm k}^4}{15{\rm c^2h^3}}T^4\]
this means that we have found an alternative expression for Stefan's
constant in terms of fundamental quantities:
\[\sigma = \frac{2\pi^5{\rm k}^4}{15{\rm c^2h^3}}\]

\subsection{The total absorbed radiation at Earths surface}

With the current interest in alternative energies, the question of
how much energy is available from the sun at earths surface becomes
very important for any application that aims at taking advantage of
this clean source of energy.

\begin{center}
  \begin{figure}
    \centerline{\includegraphics[width=.9\textwidth]{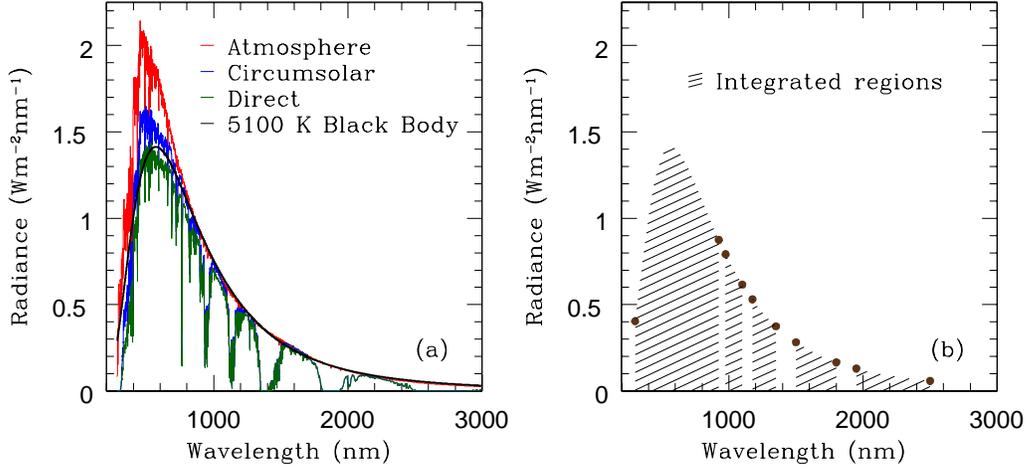}}
    \caption[Spectra of the sun]{(a) Spectra of the sun: red, as it
      reaches the atmosphere; blue, direct beam plus circumsolar
      component in a disk 2.5 degrees around the sun; green, direct
      beam. In black we plot the corresponding curve of a 5100K black
      body. (b) Approximation of filtered, direct beam, spectra as it
      reaches the surface of the earth.} 
    \label{spec}       
  \end{figure}
\end{center}

As light from the sun reaches the earth, the atmosphere blocks out
considerable parts of the spectrum, in figure \ref{spec} we have
plotted the amount of radiation which eventually reaches the surface
of the earth on a given day, in the southwest USA \cite{solardata}.
In Figure \ref{spec} (a) we have plotted in red the
standard spectra of the sun as it reaches the atmosphere, while the
other two lines show the direct components that reach the earth
surface, in blue we show the direct beam from the sun plus the
circumsolar component in a disk 2.5 degrees around the sun, while the
green is just the direct beam radiation. (although this plot is ours,
it is intended to reproduce that shown in the web page mentioned
above). We can consider this green line to indicate the total
light available to us for collection by ground systems. We also plot a
black body curve adjusted to a temperature of 5100K, shown in black,
which we use to model the direct beam component. 
As we can see, the radiation is highly filtered by the atmosphere, so
not all wavelengths reach the Earths surface. In Figure \ref{spec} (b)
we have reproduced the black body fit at 5100K as above, but have now
removed part of the spectrum which is naturally filtered. The
wavelength intervals (measured in nm) to be integrated are thus:
[300.,925.], [975., 1100.], [1175.,1350.], [1500.,1800.] and
[1950.,2500.]. To calculate the total radiance available, we will use
a modified version of (\ref{dens1}) and the relation between spectral
radiance and energy density, since we may add each contribution, we
can write the total radiance received at the surface:

\[
R_T=\frac{c}{4}\beta\sum_i\left[F(\chi_{b_i})-F(\chi_{a_i})
  \right]
\]

\noindent In this last expression, $\chi_{a_i}$ represent the
frequencies associated to the initial part of each interval: $\{300.,
975.,1175.,1500.,1950.\}$, while $\chi_{b_i}$ represent the final
part: $\{925., 1100.,1350.,1800.,2500.\}$. Using (\ref{Ffinal}) with
the method developed in the past section and a 
suitable method for numerical integration, we
can find the integrated radiance in each of these intervals and thus
the total available radiation at the earths surface. 

In table \ref{tableres} we present the value of F for the different
wavelengths chosen. If we recall that our temperature parameter
$\beta$ is determined by the black body fit to 5100K, then
$\beta=2.36289\times 10^7$Wm$^{-2}$,
and this leads us to a total radiation equivalent to:

\begin{equation}
  R_T=1.24402\times 10^8 {\rm Wm}^{-2}
  \label{result}
\end{equation}

\begin{table}[h]
  \begin{center}
\begin{tabular}{|c|c|c|c|}
  \hline
  $\lambda_a$ (nm) & F & $\lambda_b$ (nm) & F \\
  \hline
   300. & 0.09554 &  925. & 3.87127 \\
   975. & 4.09277 & 1100. & 4.55882 \\
  1175. & 4.78702 & 1350. & 5.20358 \\
  1500. & 5.46369 & 1800. & 5.81167 \\
  1950. & 5.92909 & 2500. & 6.1876 \\
  \hline  
\end{tabular}
\caption[Radiance results]{Auxiliary function, $F$, values for the
  wavelengths of interest defined in Figure \ref{spec}}
\label{tableres}
\end{center}
\end{table}


\section{Conclusions}

We have presented a review of a compact form to calculate the total
radiance output from a black body between two
wavelengths. Considering modern computational techniques and power,
the expressions presented are relatively easy to manage and can lead to
quick results.

These results may allow better estimates of the available power for
radiance capturing experiments. If we were to model the total
radiation at the earths surface by the simple fit to a 5100K black body
without considering that there are filtered regions and simply
integrate through the whole wavelength range (300 nm to 2500 nm) then the
calcultaed radiance obtained in equation \ref{result} would be only
slightly more that 87\% of the total radiance calculated. Therefore,
we beleave that a technique such as the one presented in this work are of
special importance if we are to model correctly the available energy.

\section{Acknowledgements}
The authors would like to acknowledge the generosity of Dr. Christiana
Honsberg  and Dr. Stuart Bowden at the Solar Power Labs ASU
(http://pv.asu.edu/), for making the solar spectra available in an
easy to use format at www.pveducation.org. D. Santana and O. Salcido
acknowledge the generous support of CONACyT through a grant for
postgraduate training. This project was partially supported by a grant
from the Divisi\'on de Ciencias Exactas y Naturales of the Universidad
de Sonora, grant no: USO315001752.

\label{concls}
%
%


\section*{References}
%
 \bibliographystyle{iopart-num}
 \bibliography{BBRadCalSalSan}

\end{document}